\documentclass{article}

\usepackage{PRIMEarxiv}

\usepackage[utf8]{inputenc} 
\usepackage[T1]{fontenc}    
\usepackage{hyperref}       
\usepackage{url}            
\usepackage{booktabs}       
\usepackage{amsfonts}       
\usepackage{nicefrac}       
\usepackage{microtype}      
\usepackage{lipsum}
\usepackage{fancyhdr}       
\usepackage{graphicx}       
\graphicspath{{media/}}     
\usepackage{amsmath,amssymb,amsfonts}
\title{Classical Capacity of Arbitrarily Distributed Noisy Quantum Channels
\thanks{\textit{This material is based upon work supported by Science Foundation Ireland (SFI) and is co-funded under the European Regional Development Fund under Grant Numbers 13/RC/2077 and 13/RC/2077-P2. \\
\copyright 2023 IEEE. Personal use of this material is permitted. Permission from IEEE must be obtained for all other uses, in any current or future media, including reprinting/republishing this material for advertising or promotional purposes, creating new collective works, for resale or redistribution to servers or lists, or reuse of any copyrighted component of this work in other works.}} 
}

\author{\uppercase{I. Dey}, \uppercase{H. Siljak} and \uppercase{N. Marchetti}}

\begin{document}
\maketitle

\begin{abstract}
With the rapid deployment of quantum computers and quantum satellites, there is a pressing need to design and deploy quantum and hybrid classical-quantum networks capable of exchanging classical information. In this context, we conduct the foundational study on the impact of a mixture of classical and quantum noise on an arbitrary quantum channel carrying classical information. The rationale behind considering such mixed noise is that quantum noise can arise from different entanglement and discord in quantum transmission scenarios, like different memories and repeater technologies, while classical noise can arise from the coexistence with the classical signal. Towards this end, we derive the distribution of the mixed noise from a classical system's perspective, and formulate the achievable channel capacity over an arbitrary distributed quantum channel in presence of the mixed noise. Numerical results demonstrate that capacity increases with the increase in the number of photons per usage. 
\end{abstract}


\section{Introduction}

The classical capacity of a quantum channel is the maximum possible data rate at which classical information can be transmitted through a quantum channel. On the other hand, the quantum capacity of a quantum channel is the maximum data rate at which quantum information can be transferred over a quantum channel. In the former case, only the classical environment is considered disregarding any quantum effects; in the latter scenario, quantum effects are considered without any classical world impairments. However, in a realistic communication scenario with quantum links carrying classical information, both classical and quantum channel effects and uncertainties will be encountered \cite{1b}.

Classical capacity can be enhanced by exploiting pre-shared entangled states as a connecting resource between two end-nodes of a communication link. Under ideal conditions, capacity can be doubled by sharing entanglement as compared to unassisted capacity. The effect of a non-ideal (noisy) channel on the entanglement-assisted classical capacity of a quantum channel was studied in \cite{3d}. However, \cite{3d} is limited to a linear channel with Gaussian distributed input, output and classical noise without accounting for any quantum-domain uncertainties and noise. The same is the case for Shannon's bound \cite{1a} or Holevo bound \cite{3b} or Lloyd-Shor-Devetak (LSD) bound \cite{3c}, where \cite{1a} and \cite{3b} concentrate on classical capacity of quantum channels with Gaussian input states and Gaussian classical noise and \cite{3c} accounts for quantum capacity of a Gaussian quantum channel.

To realize the full potential of quantum communication systems when used to transfer classical information, we need to encode classical information into a set of quantum states at the input of the quantum channel, and decode those quantum states back to classical information through measurement at the output of the quantum channel. \emph{In this paper, we make the first ever attempt to answer the generic question of what the capacity of such a channel will be. We assume an arbitrarily distributed quantum channel, plagued with both classical and quantum noise.} We relate the communication theorist’s concept of classical noise and uncertainties with the physicist’s description of quantum field noise, where the relationship is additive. Using this additive relationship, we derive the hybrid noise model mathematically and calculate the achievable capacity of an arbitrary quantum channel carrying classical information affected by joint quantum-classical noise and uncertainties.

Channel capacity can be calculated as the maximum number of distinguishable longitudinal modes of a propagating quantum field, that can arrive at the output of the channel over a certain duration of time and bandwidth of
operation. The longitudinal modes are essentially re-scaled electromagnetic wave functions of a harmonic oscillator, multiplied by phase factors arising from a specific geometry of the beam of quantum particles (or photons). The amplitude of a focused paraxial beam (classical electromagnetic equivalent description of a beam of photons) corresponds to the amplitude of a quantum harmonic oscillator, where the intensity of the classical electromagnetic beam corresponds to the probability distribution of the quantum oscillator. Therefore in-spite of their difference in spatial and temporal scales, it is possible to draw an analogy between the paraxial optical wave equation and the stationary and time-dependent Schrodinger equations \cite{3} guiding the propagation of a quantum field and pre-shared entangled particles, respectively, as is done in \cite{4}, \cite{5}. 
   
The paraxial approximation to the scalar Helmholtz equation \cite{3} corresponds to the Schrodinger equation, and the Guoy-phase of classical wave optics \cite{5b} corresponds to the time-coordinate of the quantum harmonic oscillator. Therefore, it is possible to map the equivalence of the qualitative behavior of the field and intensity distribution of the classical beam, with the amplitude and the probability distributions of quantum harmonic oscillators and vice-versa. Stemming from these observations, \emph{we bring in the equivalence between quantum and classical noise (uncertainties) only in terms of their probability distributions, instead of the evolution of the two noise processes in spatial and temporal scales}.

The rest of the paper is organized as the following. Section~II summarizes different channel capacity formulations for single Gaussian quantum communication channels. Section~III derives classical capacity of an arbitrary noisy quantum channel from a classical perspective. Section~IV demonstrates how the capacity of the noisy channel changes with the number of photons used per link to exchange classical information. Section~V provides concluding remarks on our work.

\section{Single Gaussian Channel Classical Capacity}\label{S2}

Owing to the huge randomness in the quantum world and the large variety of channels, modes and conditions that can exist, it has always been very difficult for scientists to pin down on a single generalized equation for the capacity $\mathcal{C}$ of a quantum channel carrying classical information. The best way forward will be to develop a physical understanding of the single linear quantum channel capacity before proceeding with a formal development. We can start with the Shannon's theorem which quantifies $\mathcal{C}$ for a single classical communication channel plagued with additive white Gaussian noise (AWGN) given as,
\begin{equation}\label{eq1}
    \mathcal{C}_c = B \log_2\bigg(1 + \frac{S}{N_0 B}\bigg)~~\text{bits/sec}
\end{equation}
where $S$ is the signal power, $B$ is the channel bandwidth (in Hz) and $N_0$ is the (white) power spectrum of channel noise (i.e. noise power per Hz). It is worth-mentioning here that the noise is band-limited and the input waveform is constrained to an ideal lowpass or bandpass bandwidth $B$. The input and output waveforms are time-limited. The ratio $S/N_0$ is referred to as the signal-to-noise ratio (SNR) or channel SNR (CSNR) and is an important metric for classical communication system. The SNR depends not only on the plaguing noise but also on the signal power available at either end of the channel, where the output signal power is again controlled by the channel properties. The Shannon bound expressed in (\ref{eq1}) is valid only if the probability density function (PDF) of the input and output waveforms are Gaussian. Intuitively from (\ref{eq1}), we can say that there is no limit on $\mathcal{C}$ for a classical channel as there is no bound on the SNR of a classical signal. SNR can only be bounded by the channel properties ($h$) and the channel noise ($n$).\\

The next thing to discuss is how to calculate $\mathcal{C}$ for a single quantum channel heuristically. The first step would be to replace the AWGN with quantum mechanical noise, disregarding the effect of classical information passing through the channel. Unlike classical signals, the energy contained within the quantum field is bounded and it follows the Einstein's relation, $E = \hbar f$ where $E$ is the energy contained in the quantum field (quantum-mechanical noise energy), $\hbar$ is the Planck's constant and $f$ is the frequency of operation. It is worth-mentioning here that our signal is classical, and the noise introduced by the quantum channel is quantum-mechanical in nature. \\

Since the energy content within a quantum field is bounded, the spectral density of quantum-mechanical noise is lower bounded by $N_0 \geq \hbar f$. Using Shannon's theorem as the underlying concept, we can write the quantum capacity of a linear quantum channel as,
\begin{equation}\label{eq2}
    \mathcal{C}_q = B \log_2\bigg(1 + \frac{S}{\hbar f B}\bigg)~~\text{bits/sec}.
\end{equation}
It is noteworthy here, that though bounded, quantum noise is neither additive nor Gaussian or white. The power spectral density (PSD) is bounded by the product of the Planck's constant and the frequency of operation.\\

Let us go back again to the physical interpretation of (\ref{eq2}). Quantum particles (photons or bosons) exhibit wave-particle duality. That means the bosons used in a quantum channel to transport classical information can exist in both the wave and the packet regimes, where for the wave-like regime $S/\hbar f B >> 1$ (many bosons/signals per longitudinal mode) and for the particle-like regime $S/\hbar f B << 1$ (many longitudinal modes per signal). Therefore, for the wave-like regime, (\ref{eq2}) can be modified to express,
\begin{align}\label{eq3}
    \mathcal{C}_{q-w} &= \big(\text{rate of transmission of modes}\big) \times\log_2 \big(\text{maximum number of distinguishable} \text{states per mode}\big) \nonumber\\
    &= B \log_2\bigg(\frac{\gamma S}{\hbar f B}\bigg)~~\text{bits/sec}.
\end{align}
where $\gamma$ is a constant of order unity. For the particle-like regime, (\ref{eq2}) can be modified as,
\begin{align}\label{eq4}
    \mathcal{C}_{q-p} &= \big(\text{rate of transmission of bosons}\big)\times\log_2 \big(\text{maximum number of distinguishable} \text{modes per quantum state}\big)\nonumber\\
    &= \frac{S}{\hbar f} \log_2\bigg(\frac{\hbar f B}{S}\bigg)~~\text{bits/sec}.
\end{align}
An intuitive conclusion that can be made from (\ref{eq4}) is that if $\hbar \to 0$, the particle-like behavior fades away, leaving only the classical wave-like behavior. We will revisit (\ref{eq3}) and (\ref{eq4}) a little bit later.

If the noise, input and output waveforms are assumed to be Gaussian distributed, the classical capacity over a Gaussian quantum channel carrying classical information without entanglement is upper bounded by the Holevo bound \cite{3b}. The Holevo bound is expressed as,
\begin{align}\label{eq4a}
   &\mathcal{C}_{c-q} = \bigg(1 + \frac{N(f)B + S\alpha}{\hbar f B}\bigg)\log_2\bigg(1 + \frac{N(f)B + S\alpha}{\hbar f B}\bigg) - \bigg(\frac{N(f)B + S\alpha}{\hbar f B}\bigg)\log_2\bigg(\frac{N(f)B + S\alpha}{\hbar f B}\bigg) + \bigg(\frac{N(f)}{\hbar f}\bigg)\nonumber\\
    &~\times\log_2\bigg(\frac{N(f)}{\hbar f}\bigg) - \bigg(1 + \frac{N(f)B}{\hbar f}\bigg)\log_2\bigg(1 + \frac{N(f)B}{\hbar f}\bigg)
\end{align}
where $N(f)/(\hbar f)$ is the noise PSD and $\alpha$ is the amplification factor of the quantum channel. The quantum capacity of a class of Gaussian quantum channels with input Gaussian states can be expressed as \cite{3c},
\begin{align}\label{eq4b}
    \mathcal{C}_{q-q} =& \max\{0, \log_2|\alpha| - \log_2|1 - \alpha|\}.
\end{align}
If the transmitter and the receiver pre-share entanglement, then the classical capacity of a Gaussian quantum channel modifies to \cite{3d}, 
\begin{align}\label{eq4c}
    \mathcal{C}_{q-ce} =&~g(S) + g(S') - g\bigg(\frac{D + S' - S - 1}{2}\bigg) - g\bigg(\frac{D - S' + S - 1}{2}\bigg)
\end{align}
where $S'$ is the average output signal energy, $D$ corresponds to the displacement operator capable of extracting magnitude and associated phase from the complex number representation and $D = \sqrt{(S + S' + 1)^2 - 4\alpha^2 S(S + 1)}$,~$g(S)$ is the entropy of the input and $g(S')$ is the entropy of the output. 

\section{Single Arbitrary Channel Classical Capacity}\label{S3}

Equations (\ref{eq1}) and (\ref{eq2}) express the upper bound for channel capacity when classical information is sent over a Gaussian classical channel, and when quantum information is sent over a Gaussian quantum channel, respectively; considering that both the channels are linear and the PDFs of the input and output waveforms are Gaussian. Now a bigger question (and mostly unsolved) is what will happen if we want to send classical information over a quantum channel. If we assume that the quantum channel is linear and the PDFs of the input and output waveforms of the classical channel are Gaussian, then heuristically we can write,
\begin{align}\label{eq5}
    \mathcal{C}_{cqc} &= B \log_2\bigg(1 + \frac{S}{B N_{xy}(f)}\bigg)~~\text{bits/sec}.
\end{align}
where $N_{xy}(f)$ is the Cross-PSD of classical additive white Gaussian noise $X$ and quantum mechanical noise $Y$. Now a Gaussian PDF for both the input and output waveforms of the classical information means either a coherent state or a quadrature-squeezed channel. For coherent channels, classical information at the input is contained in the complex field, i.e. both the (real) amplitude and phase. At the output, information can be retrieved by measuring both the real and imaginary parts. Such a measurement will involve non-commuting observables, if received at the end of a quantum channel. However, such a measurement will only be degraded by quantum noise and conventional quantum uncertainties of the observables. Therefore coherent channels never represent a quantum channel carrying classical information in practice. Observables over a practical quantum channel will be degraded by both quantum and classical noise and both classical and quantum uncertainties (classical uncertainties include random arbitrary phase-space displacement).

Relative to the coherent state channel, the quadrature-squeezed state channel model is even more constrained. Information is transmitted over the coherent excitation of only one of the components in the complex plane (either amplitude (real) or phase (imaginary)). Information is read out only by measuring one of the quadrature states and can be detected through a homodyne receiver. Quadrature-squeezed state channels are highly impractical as a model, though they are easy to handle in terms of formulating error-correction, transmission and resource management algorithms. Quantum uncertainties will be contained in only one quadrature component.

If the measurement of the observables is degraded by both classical and quantum noise, the channel capacity can heuristically be given by (\ref{eq5}). Now if the input and output waveforms have PDFs that follow arbitrary distributions, the resultant measurement of the observable will be crippled by classical and quantum uncertainties. Let us consider that the arbitrarily distributed channel envelope is represented by the random variable $Z$. The channel SNR can then be expressed as, $\gamma = Z^2\frac{S}{N_{xy}(f)}$. For a continuous-input continuous-output (memoryless) linear arbitrarily distributed communication channel, the capacity can be derived in terms of,
\begin{align}\label{eq6}
    \mathcal{C}_{cqc} = \mathrm{E}_{\gamma}\big[B \log_2 (1 + \gamma)\big]~~\text{bits/symbol}.
\end{align}
where the expectation is taken over $\gamma$. Equation (\ref{eq6}) can be rewritten as,
\begin{align}\label{eq7}
    \mathcal{C}_{cqc} &= \int_0^{\infty} B \log_2(1 + \gamma) f_{\gamma}(\gamma) \mathrm{d}\gamma = \int_0^{\infty} B \log_2 \bigg(1 + \frac{z^2 S}{N_{xy}(f)}\bigg) f_{\gamma}\bigg(\frac{z^2 S}{N_{xy}(f)}\bigg) \mathrm{d}\gamma.
\end{align}
Using transformation of variables we can calculate,
\begin{align}\label{eq8}
    \mathrm{d}\gamma = \mathrm{d}\bigg\{\frac{z^2 S}{N_{xy}(f)}\bigg\} = \frac{2zS}{N_{xy}(f)} \mathrm{d}z.
\end{align}
Putting (\ref{eq8}) in (\ref{eq7}), we can express the generalized capacity as,
\begin{align}\label{eq9}
    &\mathcal{C}_{cqc} = \int_0^{\infty} B \log_2 \bigg(1 + \frac{z^2 S}{N_{xy}(f)}\bigg) f_Z\bigg(\frac{z^2 S}{N_{xy}(f)}\bigg) \frac{2zS}{N_{xy}(f)}\mathrm{d}z = \frac{2BS}{N_{xy}(f)} \int_0^{\infty} z \log_2 \bigg(1 + \frac{z^2 S}{N_{xy}(f)}\bigg) f_Z\bigg(\frac{z^2 S}{N_{xy}(f)}\bigg)\mathrm{d}z
\end{align}
\begin{figure*}[t]
\normalsize
\begin{align}\label{eq20}
   \text{Var}[\mathcal{N}] = \sum_{i = 1}^M \sum_{f = 0}^{KTn_i/\hbar\tau} &\bigg[\frac{KTn_i^2e^{-\frac{K^2T^2n_i^4}{2\hbar^2\tau^2N_0B}}}{p!\hbar\tau\sqrt{2\pi N_0B}} \bigg(\frac{\hbar f\tau}{KT}\bigg)^p e^{\frac{n_i^2f}{N_0B} + \frac{\hbar\tau f}{KT} - \frac{(\hbar f\tau)^2}{2N_0BK^2T^2}} \nonumber\\
   &- \frac{K^2T^2n_i^2e^{-\frac{K^2T^2n_i^2}{\hbar^2\tau^2N_0B}}}{p!^2\hbar^2\tau^2 2\pi N_0B} \bigg(\frac{\hbar f\tau}{KT}\bigg)^{2p} e^{\frac{2n_if}{N_0B} + \frac{2\hbar\tau f}{KT} - \frac{(\hbar f\tau)^2}{N_0BK^2T^2}}\bigg]
\end{align}
\hrulefill
\end{figure*}
In order to solve (\ref{eq9}), the first step is to derive an expression for the Cross-PSD $N_{xy}(f)$, which is the cross PSD of the classical noise $X$ which is Gaussian distributed with zero mean and variance $N_0B$, and the quantum mechanical noise $Y$ which can be initially assumed to be Poisson distributed. It is worth-mentioning here that, in general, if photons are considered to be the carrier of quantum information, quantum noise can be assumed to be Poisson distributed \cite{33b}. In that case, the quantum intensity noise can be expressed in terms of its probability mass function (PMF),
\begin{align}\label{eq10}
    f_Y(y) = \frac{y^p}{p!}\,e^{-y}
\end{align}
where $Y = \frac{\hbar f\tau}{KT}$, $p$ is the number of photons travelling over a time interval $0 \leq t \leq \tau$, $f$ is the frequency at which photons are emitted or transmitted, $K$ is the Boltzmann's constant, and $T$ is the temperature of operation. The PDF of the classical AWGN on the other hand can be given by,
\begin{align}\label{eq11}
    f_X(x) = \frac{1}{\sqrt{2\pi N_0 B}}\,e^{-\frac{x^2}{2 N_0 B}}.
\end{align}

To compute $N_{xy}(f)$, we need to find the variance of the joint variable representing the combined classical and quantum noise. We know that classical noise is additive. If quantum noise is also assumed to be additive, the joint noise variable can be given by,
\begin{align}\label{eq12}
   \mathcal{N} := X + Y.
\end{align}
\begin{figure*}[t]
\normalsize
\begin{align}\label{eq21}
   N_{xy}(f) = N_{\mathcal{N}}(f) = \int_{-\infty}^{\infty} \text{Var}[\mathcal{N}] e^{-j2\pi ft}\mathrm{d}t &= \sum_{i = 1}^M \sum_{f = 0}^{KTn_i/\hbar\tau} \bigg[\frac{KTn_i^2e^{-\frac{K^2T^2n_i^4}{2\hbar^2\tau^2N_0B}}}{p!\hbar\tau\sqrt{2\pi N_0B}} \bigg(\frac{\hbar f\tau}{KT}\bigg)^p \frac{e^{\frac{n_i^2f}{N_0B} + \frac{\hbar\tau f}{KT} - \frac{(\hbar f\tau)^2}{2N_0BK^2T^2}}}{j2\pi f} \nonumber\\
   &- \frac{K^2T^2n_i^2e^{-\frac{K^2T^2n_i^2}{\hbar^2\tau^2N_0B}}}{p!^2\hbar^2\tau^2 2\pi N_0B} \bigg(\frac{\hbar f\tau}{KT}\bigg)^{2p} \frac{e^{\frac{2n_if}{N_0B} + \frac{2\hbar\tau f}{KT} - \frac{(\hbar f\tau)^2}{N_0BK^2T^2}}}{j2\pi f}\bigg]
\end{align}
\hrulefill
\end{figure*}
Noise can be modelled as the complex interaction between each particle of the medium and the rest of the system. Noise within any medium is a consequence of the large number of statistically independent interactions of the particles with the wave, or particles (molecules/quantum particles) that are propagating through the medium. Using Planck's radiation law and the concept of linear Stoke's friction, any kind of noise can be expressed as an additive phenomenon, $A(= a_1 + a_2 + \dotso + a_j + \dotso + a_N)$, where, $a_j$ can be derived from a distribution with zero mean and variance equal to a random number derived from the fluctuation-dissipation theorem \cite{34a}. Possible distributions from which $a_j$ can be derived are $\alpha$- stable Lévy type \cite{35}, Laplace distribution \cite{36}, Gaussian distribution, or Poisson distribution. 
\begin{figure*}[t]
\begin{minipage}{0.45\linewidth}
\centering
\includegraphics[width=0.99\textwidth]{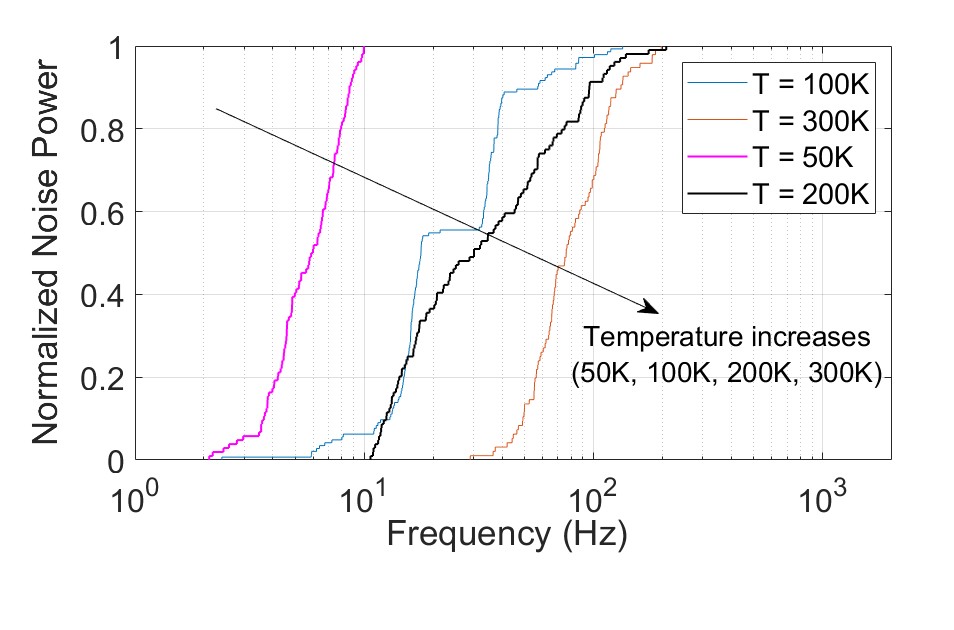}
\vspace*{-10mm}
\caption{Variation of the normalized noise power with frequency of operation where the classical noise bandwidth $B = 10$KHz and 1000 photons are traversing in unit time.}
\label{FIG1}
\end{minipage}
\hspace{0.2cm}
\begin{minipage}{0.5\linewidth}
\centering
\includegraphics[width=1.05\textwidth]{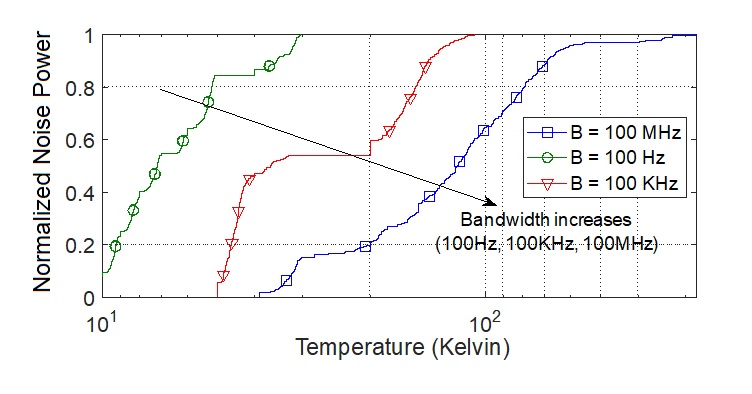}
\vspace*{-6mm}
\caption{Variation of the normalized noise power with temperature $T$ in degree Kelvin where $f = 100$ MHz and 1000 photons are traversing in unit time.}
\label{FIG2}
\end{minipage}
\hspace{0.3cm}
\vspace*{-3mm}
\end{figure*}

Combination of classical and quantum noise can take any form, such as additive $\mathcal{N} = X + Y$, or multiplicative $\mathcal{N} = X * Y$, or classical noise being a function of quantum noise, $\mathcal{N} = X(Y)$. In this paper, we will consider the joint classical-quantum noise to follow an additive combination. In order to calculate the joint PDF, $f_{\mathcal{N}}(n)$, we have to combine a discrete random variable with a continuous random variable. We can start by finding the cumulative distribution function (CDF) as,
\begin{align}\label{eq13}
   F_{\mathcal{N}}(n) &= P(\mathcal{N} \leq n) = P(X + Y \leq \mathcal{N}) = \sum_y P(X \leq \mathcal{N} - Y)P_Y(y) = \sum_y F_X(n - y)P_Y(y)
\end{align}
The PDF can be derived from (\ref{eq13}) as,
\begin{align}\label{eq14}
   f_{\mathcal{N}}(n) &= \frac{\mathrm{d}}{\mathrm{d}n} F_{\mathcal{N}}(n) = \frac{\mathrm{d}}{\mathrm{d}n} \sum_y F_X(n - y)P_Y(y) = \sum_y \frac{\mathrm{d}}{\mathrm{d}n} F_X(n - y)P_Y(y) = \sum_y f_X(n - y)P_Y(y).
\end{align}
where $\mathcal{N}$ is a hybrid continuous discrete variable. Putting (\ref{eq10}) and (\ref{eq11}) back in (\ref{eq14}), we can express the PDF of the joint noise variable as,
\begin{align}\label{eq15}
   f_{\mathcal{N}}(n) &= \sum_{y = 0}^n \frac{e^{-n^2/2N_0B}}{p!\sqrt{2\pi N_0B}} y^p e^{\frac{ny}{N_0B} + y - \frac{y^2}{2N_0B}} = \sum_{y = 0}^n \frac{e^{-n^2/2N_0B}}{p!\sqrt{2\pi N_0B}} \bigg(\frac{\hbar f\tau}{KT}\bigg)^p e^{\big(\frac{n}{N_0B} + 1\big)\frac{\hbar f\tau}{KT} - \frac{(\hbar f\tau)^2}{2N_0BK^2T^2}}.
\end{align}
Using the change of variables, we can express (\ref{eq15}) as,
\begin{align}\label{eq16}
   f_{\mathcal{N}}(n) = \sum_{f = 0}^{KTn/\hbar\tau} \frac{e^{-\frac{K^2T^2n^2}{2\hbar^2\tau^2N_0B}}}{p!\sqrt{2\pi N_0B}} \bigg(\frac{\hbar f\tau}{KT}\bigg)^p e^{\frac{nf}{N_0B} + \frac{\hbar\tau f}{KT} - \frac{(\hbar f\tau)^2}{2N_0BK^2T^2}}.
\end{align}
Using the initial PDF expression of mixed noise in (\ref{eq15}), we can derive the mean, second moment and variance of $\mathcal{N}$ as,
\begin{align}\label{eq17}
   \mathrm{E}[\mathcal{N}] &= \sum_{i = 1}^M n_i P(\mathcal{N} = n_i) = \sum_{i = 1}^M \sum_{y = 0}^{n_i} \frac{n_i e^{-n_i^2/2N_0B}}{p!\sqrt{2\pi N_0B}} y^p e^{\frac{n_iy}{N_0B} + y - \frac{y^2}{2N_0B}}
\end{align}
\begin{align}\label{eq18}
   \mathrm{E}[\mathcal{N}^2] = \sum_{i = 1}^M \sum_{y = 0}^{n_i} \frac{n_i^2 e^{-n_i^4/2N_0B}}{p!\sqrt{2\pi N_0B}} y^p e^{\frac{n_i^2 y}{N_0B} + y - \frac{y^2}{2N_0B}}
\end{align}
\begin{align}\label{eq19}
   \text{Var}[\mathcal{N}] &= \sum_{i = 1}^M \sum_{y = 0}^{n_i} \bigg[\frac{n_i^2 e^{-n_i^4/2N_0B}}{p!\sqrt{2\pi N_0B}} y^p e^{\frac{n_i^2 y}{N_0B} + y - \frac{y^2}{2N_0B}} - \frac{n_i^2 e^{-n_i^2/N_0B}}{p!^2 2\pi N_0B} y^{2p} e^{\frac{2n_i y}{N_0B} + 2y - \frac{y^2}{N_0B}}\bigg]
\end{align}
and the variance of $\mathcal{N}$ can be expressed in terms of the frequency of operation $f$ as in (\ref{eq20}), by a change of variable. The PSD of $\mathcal{N}$ can be calculated by applying the Fourier Transform to $\text{Var}[\mathcal{N}]$, to obtain (\ref{eq21}) where $p$ is the average number of photons per use, $p = S/\hbar f B$. If $p$ is the average number of photons per second per Hz, the average transmission rate is $Bp$, and hence the average input power to the channel is given by, $S = B\hbar f p$ and the average number of photons per use $p = S/\hbar f B$ is the most crucial dimensionless quantity. The capacity with Gaussian input and output therefore can be obtained by putting (\ref{eq21}) back in (\ref{eq5}). 
\begin{figure}[t]
\centering
\vspace*{-6mm}
\includegraphics[width=0.99\linewidth]{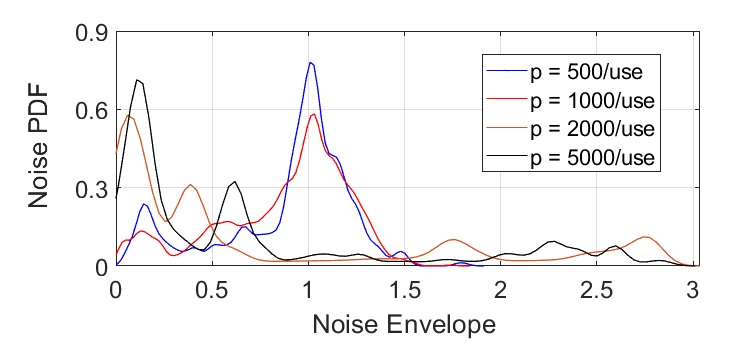}
\caption{PDF of the noise envelope with varying photon numbers traversing per use with $B = 10$KHz, $f = 100$MHz and $T =290$K, i.e. room temperature.}
\label{FIG3}
\vspace*{-6mm}
\end{figure}

\section{Numerical Results and Discussions}\label{S4}

In this section, we demonstrate how our developed joint classical-quantum noise power changes with different factors affecting it, and how its behaviour shifts between the dominant classical and quantum regimes depending on the environmental parameters. Based on the noise samples generated by the PDF in (\ref{eq16}), we also portray how the upper bound on channel capacity changes when classical information is sent over an arbitrarily distributed quantum channel. We start by deciding on the number of noise samples we generate, i.e., $M = 10$. We choose a large enough $M$ so that the noise PDF exhibits similar characteristics on an average over the entire range of frequency, temperature and bandwidth of operation.

\begin{figure}[t]
\centering
\includegraphics[width=0.9\linewidth]{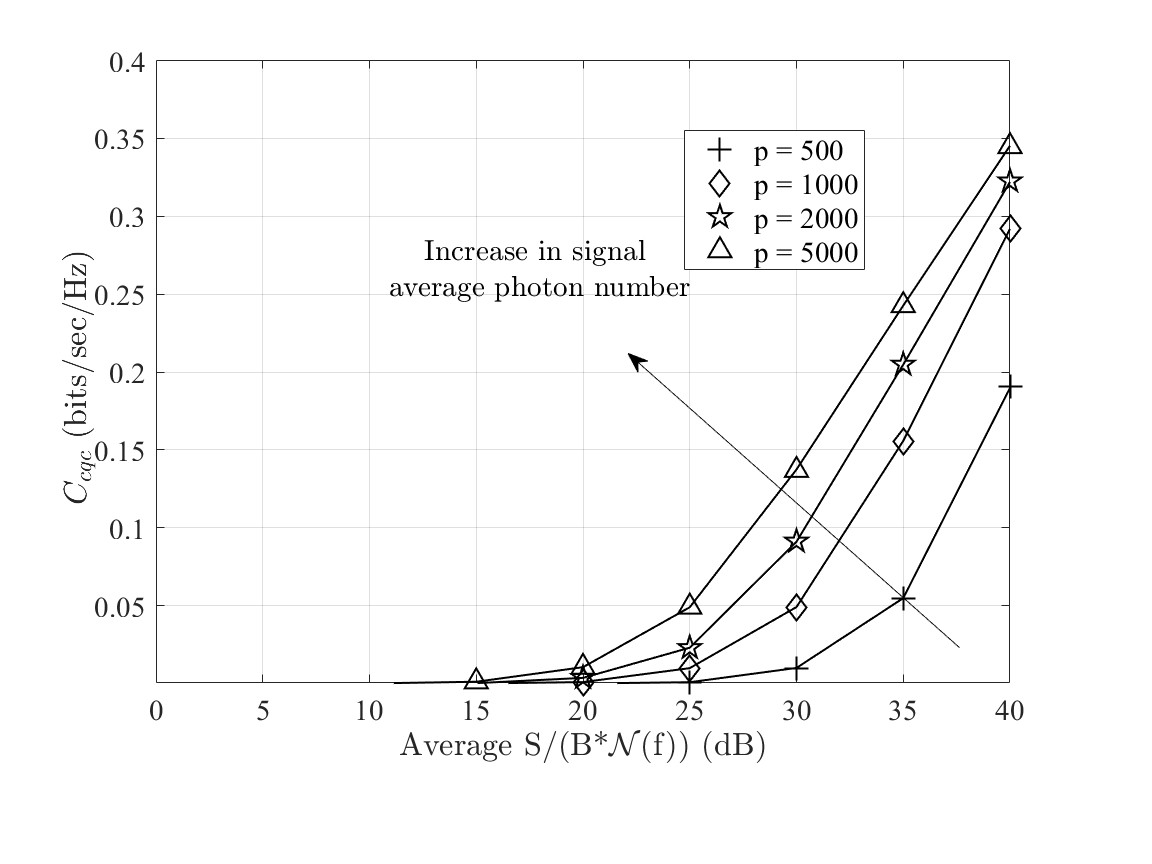}
\vspace*{-5mm}
\caption{Achievable Capacity with increasing average unit signal-to-noise per bandwidth ratio with $B = 10$KHz, $f = 100$MHz and $T =290$K, i.e. room temperature.}
\label{FIG4}
\vspace{-6mm}
\end{figure}

The first set of curves (Fig.~\ref{FIG1}) are generated by varying the normalized noise power with frequency and at different temperatures. Noise power increases with frequency; however the rate of increase slows down with the increase in temperature. The reason can be attributed to the fact that at lower temperatures, the noise PSD is the mean-squared current fluctuation per unit bandwidth. At higher temperatures, the noise samples start behaving closer to the quantum shot noise regime which becomes frequency-independent and linearly proportional to the current. 

A very similar effect is visible when noise is varied with temperature where the rate of increase in noise power shows downward trend with increase in bandwidth (refer to Fig.~\ref{FIG2}). Using the relation, $\mathcal{N}_{xy}(f) \propto \langle \partial I(f)^2 \rangle/B$ where $I(f)$ is the current fluctuation due to frequency, we can see that the noise PSD is inversely proportional to the bandwidth. This inverse relationship with bandwidth results in the slow hike in noise power with temperature increase. For both the results in Fig.~\ref{FIG1} and Fig.~\ref{FIG2}, we used a moderate number of photons per use, i.e., $p = 1000$.

The third set of results are generated by plotting the PDF of $\mathcal{N}_{xy}$. By varying the photon number per use, we generated the PDFs and we can see quite a bit of difference in characteristics as we move from a moderate $p$ to a large $p$. When $p$ is moderate, we see the noise behavior dominated by the classical Gaussian characteristics with an impulse-like nature, narrower at the top and wide at the bottom. When $p$ increases, the noise is dominated by the Poisson light-like quantum characteristics with small humps appearing at the tail. The plots in Fig.~\ref{FIG3} therefore validate the fact that the noise characteristics is highly dominated by the average number of photons per use.

With the increase in $p$, the average input power to the channel $S = B\hbar fp$ increases, which results in an increase in channel capacity per unit bandwidth, as is evident in Fig.~\ref{FIG4}. However, the channel capacity obtained in presence of both classical and quantum noise is way lower than is achievable over standard classical or quantum communication channel. This leaves us with the possibility to explore different parameters, like frequency of operation, system bandwidth, temperature, relation between classical and quantum-domain uncertainties etc., that will effect the capacity and can be tuned to optimize achievable capacity. 

Since in this paper we are looking at the quantum channel from the classical perspective, let us look at how the Shannon bound on the classical channel capacity will numerically compare with our derived capacity of a quantum channel carrying classical information. Classical Shannon's capacity can be calculated from (\ref{eq1}). That means for $S/N_0 = 30$ dB, and $B = 10$ KHz, $\mathcal{C}_c/B = \log_2(1 + S/(N_0 B)) = \log_2(1.1) \sim 0.14$ bits/sec/Hz. From Fig.~\ref{FIG4}, for $p = 1000$/use, $\mathcal{C}_{cqc}/B = 0.05$ bits/sec/Hz and for $p = 2000$/use, $\mathcal{C}_{cqc}/B = 0.095$ bits/sec/Hz. This scenario can be interpreted as the region where we can play with the number of photons that can possibly be used to carry the information. We can consider the Shannon classical capacity as the achievable upper bound which can be reached by controlling the number of photons and the signal bandwidth. 
\vspace{-3mm}
\section{Conclusions}\label{S5}

In this paper, we have studied the mixture of classical and quantum noise and uncertainties observed when classical information is exchanged over a quantum link. We have calculated the capacity of such links under the assumption that the mixture of classical and quantum noise is additive. We also assume that the probability distribution of the additive mixture is possible to derive, even if the evolution of the classical and quantum noise processes are different in spatial and temporal scales. We also demonstrated the impact of different factors on the additive noise mixture model, and the achievable capacity over an arbitrary quantum channel in presence of the mixed noise. Numerical results reveal that noise power increases with frequency and decreases with bandwidth. Temperature also plays an important role in the noise behavior, where the noise starts behaving closer to quantum shot noise with an increase in temperature. The number of photons used per link also has a considerable impact on channel capacity, where the capacity can be enhanced with the increase in the photon number per usage.

\end{document}